\documentclass[draft]{agujournal2019}
\usepackage{url} 
\usepackage{lineno}
\usepackage{soul}

\usepackage{amsmath,amssymb}
\usepackage{bm}

\draftfalse
\journalname{JGR: Earth Surface}

\begin{document}

\title{Surface Crevasse Evolution Observed Using Matched Field Processing and Source Relocation at Hansbreen, Svalbard}

\authors{Wojciech Gajek\affil{1}, Ugo Nanni\affil{2}, Ali Gholami\affil{1}, William D. Harcourt\affil{3,4}, Danni M. Pearce\affil{3,5}, Louise Steffensen Schmidt\affil{2} }

\affiliation{1}{Institute of Geophysics, Polish Academy of Sciences, Warsaw, Poland}
\affiliation{2}{Department of Geosciences, University of Oslo, Norway}
\affiliation{3}{School of Geosciences, University of Aberdeen, Aberdeen, UK}
\affiliation{4}{Interdisciplinary Institute, University of Aberdeen, Aberdeen, UK}
\affiliation{5}{Faculty of Environmental Science and Natural Resource Management, Norwegian University of Life Sciences, Oslo, Norway}

\correspondingauthor{Wojciech Gajek}{wgajek@igf.edu.pl}

\begin{keypoints}
\item Meters-scale resolution cryoseismicity mapping on Hansbreen, Svalbard, using sparse-array Matched Field Processing and icequake relocation.
\item Crevasse propagation rates (0.005–0.007 m/s) and viscous diffusion coefficients (0.47–0.55 m$^2$/s) derived from relocated surface icequakes.
\item Few-hour-long crevasse propagation explained by sustained subcritical cracking controlled by viscous stress relaxation below elastic limits.
\end{keypoints}

\begin{abstract}
Crevasses control glacier dynamics through fracture and meltwater routing, yet their propagation rates remain observationally scarce and poorly constrained across brittle-to-viscous regimes. Cryoseismology offers a powerful means to capture dynamic processes within glacial ice, with recent advances in novel processing methods like Matched Field Processing (MFP) applicable to dense seismic arrays. However, precise localisation of cryoseismic sources remains challenging in sparse or irregular seismic arrays. We propose a two-step workflow for metres-scale resolution mapping of glacial seismic activity that integrates MFP and discrete arrival times relocation under a limited instrumentation constraint.
We apply this approach to analyse seismic activity at the ice surface on the Hansbreen glacier, Svalbard. Using MFP, we detect surface icequakes and characterise meltwater noise regardless of the limited instrumentation. The relocation procedure increases the accuracy of surface icequakes localisation and reveals ongoing crevasse opening episodes. The precise locations of the icequakes allow for the estimation of the crevasse propagation rate and the determination of the diffusion coefficients of 0.47–0.55 $\mathrm{m^{2}/s}$. Based on the obtained results, we discuss brittle-to-viscous regime transfer and interpret the crevassing mechanism as sustained subcritical crack propagation, where viscous stress relaxation governs rates of orders of magnitude below elastic limits.
\end{abstract}

\section*{Plain Language Summary}
Crevasses are large cracks in glacier ice that strongly influence glacier sliding, calving, and meltwater transport. Observing how crevasses grow is difficult because the process can occur over timescales from seconds to hours and produces only weak seismic signals. These signals, however, provide a way to track crevasse activity indirectly.

In this study, we used seismic recordings from a small set of instruments on Hansbreen glacier (Svalbard) to map meltwater noise and to detect and locate icequakes associated with crevasse activity. The detected events align with crevasse patterns observed in satellite imagery, indicating that the seismic signals capture real fracture processes within the glacier.

We identify three episodes of crevasse growth, each extending up to ~190 m over several hours. The estimated opening rates are much slower than expected for brittle fracturing, which involves rapid crack propagation, and instead suggest a process controlled by viscous stress relaxation, where ice deforms gradually over time.

These findings demonstrate that methods commonly used with dense seismic arrays can also track crevasse growth and meltwater routing under limited instrumentation. Our results support improved glacier monitoring and more reliable assessments of future glacier behavior.

\section{Introduction}

Crevasses are among the key components of the glacial system, governing its mechanical and hydrological response to environmental forcing by storing and routing meltwater to drive basal sliding, and by promoting calving \cite{colgan2016glacier}. 
Crevasse initiation and growth are the result of fracture processes that operate under glacier-scale stress fields \cite{vaughan1993relating} and can culminate in catastrophic ice break-offs \cite{faillettaz2015avalanching}. Although large-scale (tens of km) rift opening can be tracked with low temporal resolution using satellite imagery \cite{fricker2002iceberg, marsh2025ocean}, the propagation rates of englacial crevassing events unfold over short timescales across spatially limited extents, making them difficult to capture using conventional glaciological methods. 
Cryoseismology has become an effective indirect method for monitoring and studying glaciers, using seismic waves generated by englacial and subglacial processes, and recorded by seismic sensors with a sub-second resolution
\cite{podolskiy2016cryoseismology,aster2017glacial}. Some cryoseismological studies were capable of estimating crevasse propagation rates \cite{neave1970icequakes,mikesell2012monitoring, taylor2019seismic, nanni2022dynamic,fichtner2025cascades}, yet these observations remain scarce as they require both good spatial and temporal coverage and resolution, whilst the reported timescales differ by five orders of magnitude (hundreds of m to a few mm per s). Such differences were interpreted as evidence for two distinct propagation mechanisms, described thoroughly by \citeA{ranganathan2025glacier}. Faster (minutes to several hours) is brittle fracturing reproduced by linear elastic fracture mechanics (LEFM) \cite{van1998fracture, nath2003subsurface, pralong2003continuum}, whilst slower (hours to months) is viscous creep governed by subcritical diffusive crack growth rate \cite{weiss2004subcritical, duddu2013numerical, hageman2024ice}. Crevasse propagation on hourly timescales—poorly constrained by field observations—falls within the ranges reported by studies employing both LEFM-based (supercritical) fracture models and subcritical creep models. These intermediate timescales likely represent transitions between subcritical creep and brittle fracturing regimes, highlighting the need for observations with improved temporal and spatial resolution to better constrain propagation mechanics and dynamics of transition between propagation regimes.

In recent years, dense seismic arrays \cite{arrowsmith2022big} stimulated advances in array-based processing methods enabling high-resolution mapping of glacial seismic sources. Among these methods, Matched Field Processing (MFP) \cite{baggeroer1988matched} demonstrated its effectiveness in resolving the frequency-dependent behaviour of seismic emissions related to meltwater routing, crevasse formation, and basal sliding \cite{sergeant2020green, nanni2021observing, nanni2022dynamic}.
\citeA{hudson2023array} discussed array techniques in a cryoseismological context and demonstrated trade-offs between dense arrays and distributed sensor networks, pointing out the impact of setup geometry for spatial resolution and location accuracy. However, reliance on a limited number of seismic sensors, often in 2D geometries, is still common, imposed by logistical constraints, transience of on-ice sites, or when reusing existing datasets \cite{chmiel2023hanging}.  
The extent to which modern processing methods can recover high-resolution seismic source locations from such deployments remains largely untested.
There is a need for identifying robust workflows that can resolve the spatio-temporal evolution of glacier seismicity with maximum accuracy under minimal logistical requirements, thereby enabling effective monitoring of glacier response to meteorological and hydrological forcing. 
We propose a two-step method for mapping glacial seismic activity that integrates source localisation techniques based on array processing and discrete arrival times under a limited instrumentation constraint. We apply this method to a representative, non-dense cryoseismological dataset from Hansbreen, a tidewater glacier in Svalbard, and investigate dynamic processes driven by surface crevassing and ice-meltwater interactions. 
Our analysis reveals a spatiotemporally variable seismic activity, predominantly located close to the ice surface. Using MFP, we successfully detect surface icequakes and characterise sources of meltwater-induced noise regardless of the limited instrumentation. Our improved network-based relocation procedure increases the accuracy of surface icequakes localisation and allows for detailed uncertainty analysis. Relocation reveals the crevasse-driven seismicity distribution and enables estimation of crevasse propagation rates during ongoing crevassing episodes, supporting the scenario of subcritical crevasse propagation on the timescale of a few hours.

\section{Study Site and Instrumentation}

\begin{figure}
    \centering
    \includegraphics[width=0.9\linewidth]{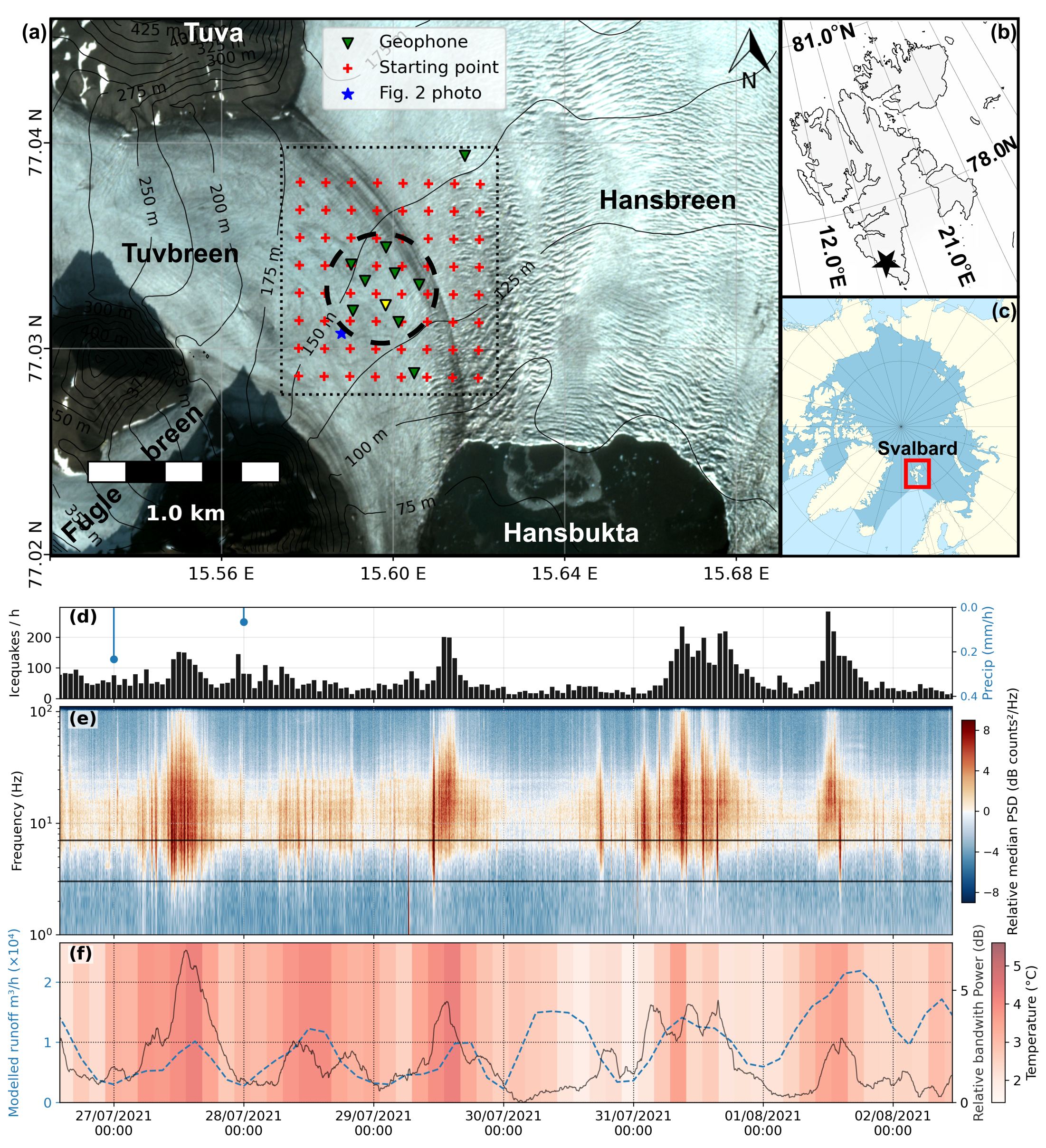}
    \caption{
Location and characteristics of the study site. 
(a) Orthophotomap showing geophone positions (triangles) at the confluence of Tuvbreen and Hansbreen in Hornsund. The yellow triangle marks the position of station B13 used to compute the spectrogram shown in panel (e). The thick, black-dashed line indicates the geophones used as a seismic array. Red crosses mark the location of starting points used in MFP computation. The dotted box represents the area shown in the Fig. \ref{fig:relocated}. The satellite photo was taken on 28/08/2021 by Planet Labs. The elevation data was sourced from the Norwegian Polar Institute and represents the situation of 2014.
(b) Location of Hornsund within the Svalbard archipelago. 
(c) Location of the Svalbard archipelago in the Arctic region. 
(d) Hourly rate of located icequakes and precipitation measured at PSP. (e) Median spectrogram calculated for station B13 (yellow triangle in (a)). Detailed procedure description and spectrograms of the remaining stations are shown in the Supporting Information (Fig. S1). 
Black lines at 3 and 7 Hz mark frequency limits used to derive median power shown in panel (f). (f) Relative median power in the bandwidth 3--7 Hz (black line), total runoff (blue dashed line) and air temperature (red-colored background) simulated for the seismic array location.
}
    \label{fig:map}
\end{figure}

The study was conducted between 26/07 and 02/08/2021 (8 days) at the confluence of Hansbreen and Tuvbreen, polythermal tidewater glaciers terminating in Hornsund Fjord, south-west Svalbard (Fig.~\ref{fig:map}), 4\,km from Polish Polar Station (PPS) Hornsund. Hansbreen is 16~km long with an average surface slope of 1.8° \cite{glazovsky1991tidewater, jania1996thermal} and a grounded calving front 1.5~km wide and up to 50~m high. It has retreated substantially since the late 19th century, including 917~m between 1992 and 2015 \cite{blaszczyk2021factors}, and is classified as a surge-type glacier \cite{osika2024hans_surge}. Tuvbreen is Hansbreen's largest tributary.

We instrumented the site with ten 5~Hz vertical SmartSolo geophones recording at 250~Hz installed directly on the ice surface about 1 km from the terminus. The configuration included an 8-element seismic array \cite{schweitzer2012array} with inner and outer ring radii of 100~m and 215~m, and two more distant auxiliary geophones that started recording on 29/07 (Fig.~\ref{fig:map}a). The measured summer-averaged surface sliding velocity at the study site was 12~m/yr, whereas the central part of Hansbreen exhibits velocities close to 100~m/yr \cite{blaszczyk2024high}. Our array was deployed \(\sim1\)~km north of the legacy sites (now stagnant and partially melted), which were occasionally instrumented till the 1990s \cite{lewandowska1964investigations,gorski2014seismic}. We revisit Hansbreen's seismicity after more than two decades without on-ice seismic deployments.

During the experiment, air temperatures at the study site oscillated diurnally above freezing (1–5°C per Copernicus Arctic Regional ReAnalysis (CARRA) reanalysis data), with very limited surface melt due to persistent cloud cover. We modeled the total runoff using the CryoGrid community model \cite{westermann2023cryogrid}, forced by meteorological variables from the CARRA reanalysis (2.5 × 2.5 km spatial resolution with 3-h temporal resolution), following the methodology of \citeA{schmidt2023meltwater}. The contributing catchment draining past the seismic array was delineated using TopoToolbox \cite{schwanghart2014topotoolbox}, and the total runoff was then calculated by summing contributions over the catchment area. 

\begin{figure}
    \centering
    \includegraphics[width=1\linewidth]{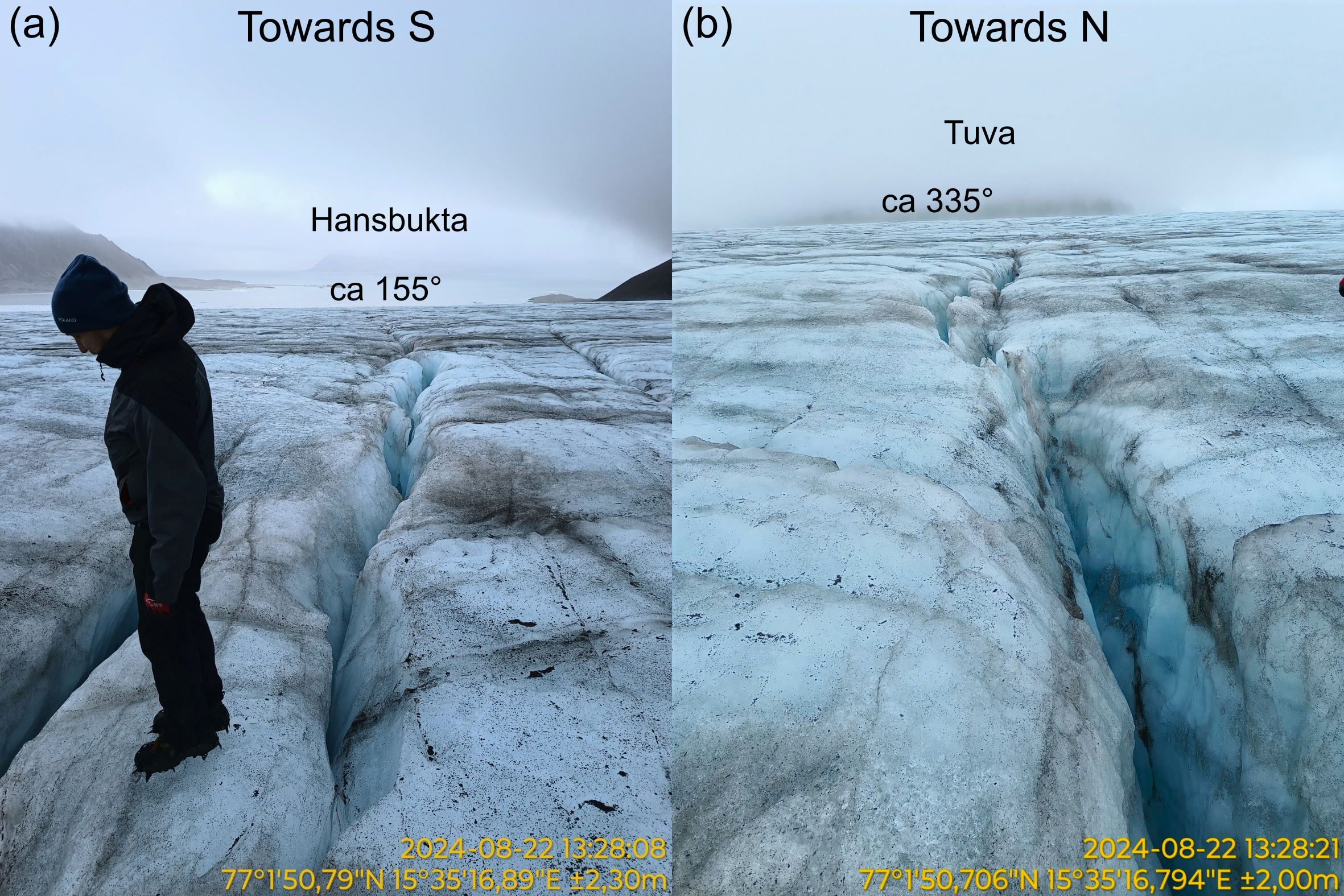}
    \caption{Photos of minor crevasses at the study site at the location of the icequake cluster shown in Fig. 8a, taken on 22/08/2024. (a) View towards the south; (b) view towards the north. The crevasse strike, estimated in the field, is $155^\circ$. }
    \label{fig:sup_crevasses_photos}
\end{figure}

Some liquid precipitation was captured on time-lapse cameras on the night of 30/07–31/07, although it was not recorded at the nearby meteorological station at PPS. Variations in diurnal temperature are reflected in modeled total runoff (Fig. \ref{fig:map}f). On-site conditions during field deployment show little to no surface melt, exposed ablated ice, and the absence of supra-glacial streams or water-filled crevasses (Fig. \ref{fig:sup_crevasses_photos}).

\section{Methods}
        To investigate the spatiotemporal patterns of cryoseismic activity and its relation to glaciological processes, we employed a combination of seismological techniques tailored to different aspects of the dataset. The following subsections detail the methods used to: (1) characterize the seismic noise and shallow subsurface velocity structure, and estimate ice thickness via analysis of surface wave dispersion; (2) detect and locate seismic sources through Matched Field Processing (MFP); (3) improve location accuracy of individual icequakes using differential arrival time inversion; and (4) assess fracture propagation patterns by analyzing temporal evolution of elongated clusters of seismicity. Each method addresses a specific step in the workflow and builds on the preceding analysis to further constrain the processes underlying the observed seismicity.

            \subsection{Shallow Velocity Structure and Noise Intensity}
            
                   Surface wave dispersion analysis was conducted using FK (frequency–wavenumber) analysis to derive frequency-dependent phase velocities of the vertical component of seismic data using stations within the array (Fig. \ref{fig:map}a). The preprocessing involved downsampling the data to 62.5 Hz (factor of 4) and 0.5--25 Hz bandpass filtering. Then, following the method proposed by \cite{kohler2004ambient}, we computed FK spectra in a sliding window for narrow frequency bands. We use a frequency-dependent window length (10 times the period) for frequencies between 0.5 and 20 Hz with a 0.5 Hz step. After processing the whole available dataset, we derived the dispersion curve by averaging the slowness values obtained from the peak location for all analysis time windows. A coherency threshold of 0.5 of the relative beam power was applied to remove the low-quality measurements when extracting the apparent velocity maxima from FK maps. Lastly, a 1D shear wave phase velocity profile was inverted from the dispersion curve using the evodcinv library \cite{luu_2021_evodcinv}.

                   To understand the evolution of noise levels over time and frequency, we compute the Power Spectral Density (PSD) of the continuous vertical ground-motion within 4-s-long windows with 25\% overlap using Welch's method \cite{welch1967use}. Then, we calculate the median value within 10-min-long windows per frequency bin. This approach minimizes the influence of discrete, highly energetic events on the PSD \cite{bartholomaus2015subglacial, preiswerk2018ambient} and allows analysis of background noise variations. 
                   
                 \subsection{Imaging the Noise Sources with Matched Field Processing}

                   Matched Field Processing (MFP) consists of localizing seismic sources in range, depth and azimuth by recursively matching the predicted and observed phase delays over a network of seismic sensors \cite{Bucker1976}. While this method has been originally developed to locate spatially well-separated sources, recent studies have also shown its potential to locate spatially distributed noise sources \cite{nanni2021observing, nanni2022dynamic}. Such a duality thus gives us the possibility to investigate both crevasses' dynamics and hydrological processes. Here, we follow the approach of \cite{nanni2022dynamic}, and conduct a systematic analysis of the phase coherence over 1-s-long time windows over our seismic array. 

For each time window, we calculate the cross-spectral density matrix $K$ by computing the discrete Fourier transform of the observed data $d$ from each station:
\begin{equation}
    K(\omega) =  \mathbf{d}(\omega) \mathbf{d}^H(\omega),
\end{equation}
where $\omega$ is the angular frequency and $H$ is the Hermetian. We then compute the synthetic data by assuming a point source $a$ located at ($x, y, z$) at a distance $r$ from the station , which generates a seismic wave propagating with a constant velocity $c$:
\begin{equation}
    d(\omega, a) = \exp(i \omega \frac{r}{c}) 
\end{equation}
To measure the difference between the measured and synthetic phase delays, we compute the Bartlett operator: 
\begin{equation}
    B(\omega,a) =\sum_{\omega} | \mathbf{d}(\omega, a)^H K(\omega)  \mathbf{d}(\omega, a)|
\end{equation}
The Bartlett operator ranges from  0 to 1, with 1 for perfect phase coherence between observed and modeled phase delays. Hereafter, we refer to this operator as the MFP output.

For each time window, we perform a search for the optimal source location and propagation velocity using a gradient-based minimization algorithm based on a downhill simplex search method \cite{lagarias1998convergence}. Such an exploration is more efficient than a gridsearch exploration, and in order to avoid local minima in the exploration of the model space, we use 64 evenly distributed starting points (see Fig. \ref{fig:map}a) located at the glacier bed 10 m a.s.l (about 120 m below the seismic array). In order to account for the possibility of simultaneously active sources, such as crevasses-induced icequakes or meltwater-induced seismic noise, we keep for each time window all results from all starting points. We then analyze our results based on both the location of the source and the associated MFP output.

The interstation distances within the array span from 92 to 417 m (the latter being the array aperture), with a mean distance of 255 m. The physical limits for MFP are fundamentally constrained by both the array aperture and the wavelengths of the recorded seismic waves, requiring that multiple wavelengths fit within the array aperture to resolve sources effectively \cite{johnson1992array, schweitzer2012array}. Hence, given our array geometry, optimal wavelengths range from 100 to 400 m. Correspondingly, for expected Rayleigh‐wave phase velocities of 1600–1700 m/s, this implies a frequency band of 4–16 Hz. Below 4 Hz (wavelengths above 400 m), the array undersamples the wavefield, yielding poor spatial resolution and ambiguous source localisation. Conversely, above 16 Hz (wavelengths below 100 m), the interstation spacing is insufficient to capture the wavefront without aliasing.

We use 4 Hz-wide frequency bands with central frequencies ranging from 4 to 16 Hz with a 2 Hz step to process continuous, downsampled to 50 Hz, data using a 1s-long sliding window with 0.5s overlap.

 \subsection{Relocation of icequakes}
                    The frequencies applicable to MFP are limited by the array geometry, resulting in geometry-limited spatial resolution. To increase the location accuracy, we apply an event relocation procedure using discrete arrival time differences derived from broader frequency range signals, as network-based methods are not constrained with signal frequency. 
                    For the relocation, we modified the method by \citeA{roux2010observation} that applies a time delay inversion via the Levenberg-Marquardt scheme \cite{tarantola1982generalized, Nocedal_2006_NO}.
                    The method automatically picks the arrival time delays of the dominant wave by applying time-domain trace cross‐correlation within a 1-s-long window centred around the median time index of traces' absolute maxima (illustrated in Fig. S2 in Supporting Information). In the original formulation of \citeA{roux2010observation}, the inter‐station time differences are inverted via the Levenberg-Marquardt optimisation algorithm for the event’s epicentral coordinates (X, Y) and the velocity, assuming the source is located at the surface.
                    In our modified workflow, we invert for X, Y, and Z (focal depth) while prescribing a fixed velocity and introducing adaptive damping that provides the method with flexibility, robustness, and efficiency by dynamically balancing the globally convergent properties of steepest descent with the rapid local convergence of Newton’s method. We present the details of relocation procedure including formulation of the inverse problem and solving algorithm in the Supporting Information, section S1. 
                    
                    We determine the velocity independently by iteratively testing a range of values and selecting the one that minimises the median misfit of all relocated events.


\subsection{Uncertainty Estimation}
To assess event location uncertainty arising from array geometry, the time delay residuals, and the inaccurate velocity model, we implemented a Monte Carlo (MC) framework \cite{sambridge2002monte}. We estimate uncertainty individually for each event based on its root-mean-square error (RMSE) between observed and modelled time delays. We add Gaussian noise scaled with RMSE to modelled time delays.
Velocity uncertainty is represented by perturbing the reference velocity in each trial with Gaussian noise proportional to our velocity uncertainty estimate. Finally, we quantify the location uncertainty as the standard deviation of recovered source coordinates across 500 trials. In contrast to \citeA{roux2010observation}, who applied a uniform timing uncertainty to all events, our approach captures event-specific variability of time delay residuals.
            
\subsection{Magnitude and Volume Change Estimation}

We compute a local surface-wave magnitude \(M_S\) to assess the relative size of recorded icequakes, following the methodology of \citeA{mikesell2012monitoring}. First, we correct Rayleigh wave amplitudes measured at multiple stations for geometric spreading using a $1/\sqrt{R}$ scaling appropriate for surface waves. Next, we analyse the corrected amplitude as a function of epicentral distance and model the attenuation-dependent amplitude decay as

\begin{equation}
A(R_i) = \frac{A_0}{\sqrt{R_i}} e^{-\alpha R_i},
\end{equation}

where \(A(R_i)\) is the observed ground displacement amplitude in microns at distance \(R_i\), \(A_0\) the source amplitude at the epicentre, and \(\alpha\) the attenuation decay constant. Finally, we extrapolate the source amplitude \(A_0\) from the linear regression of the natural logarithm of amplitude against the distance and obtain the magnitude \(M_S\) using

\begin{equation}
\label{eq:surface_magnitude}
M_S = \log_{10}(A_0),
\end{equation}

Although \(M_S\) is an observational measure, it can be linked to the physical strength of the seismic source through an empirical relationship with the seismic moment  \(M_0\) \cite{lay1995modern}:

\begin{equation}
\label{eq:m0}
\log_{10}(M_0) = 1.5M_S +16.1
\end{equation}

The \(M_0\) provides a direct measure of the strength of the seismic source, as it reflects the magnitude of the inelastic deformation that occurred during the event. For tensile cracking, the corresponding volume change dV is estimated using the relation proposed by \citeA{Mueller1991}:

\begin{equation}
\label{eq:dv}
dV = \frac{M_0}{\lambda + \tfrac{2}{3}\mu} ,
\end{equation}
where $\lambda= 5.71$ GPa and $\mu= 2.9$ GPa denote the Lamé parameters of ice, following \citeA{mikesell2012monitoring}.

\section{Results}

To study the temporal and spectral variability of seismic noise at the study site, we analyse the median PSD over time. Fig. \ref{fig:map}e shows the evolution of median noise amplitude over time and frequency for the B13 station. Observed noise variations are consistent among all the stations (Supporting Information, Fig. S1). The median PSD exhibits diurnal cycles with the highest noise levels recorded during daytime. High PSD amplitudes align with elevated air temperature, runoff (Fig. \ref{fig:map}f), and icequake rate (Fig. \ref{fig:map}d). The episodically elevated noise on 31/7 may be caused by precipitation. From the median PSD, we derive the median seismic power within 3--7 Hz bandwidth as a proxy for variations in subglacial hydraulic properties such as the pressure and capacity conditions of the subglacial hydrological system \cite{gimbert2016subglacial,nanni2020quantification}. We observe a general agreement between the median seismic power and the total runoff (Fig. \ref{fig:map}f), however, some discrepancies are observed when the modeled runoff increases at the times with no significant increase in seismic power (7/30 and 8/1). The most energetic frequency band is roughly 6--30 Hz, to which surface waves, body waves, and meltwater signatures may contribute \cite{mikesell2012monitoring, gimbert2014physical, sergeant2020green}. 

        \subsection{Seismic Sources}
           In order to better understand the nature(s) of the seismic sources,  we employ Matched Field Processing (MFP). We process the continuous data from the seismic array (8 stations) in 7 frequency bands with a 2 Hz overlap. Consequently, the results change gradually with frequency. To focus on the clearest trends, we select the three most prominent and non-overlapping frequency bands: 4--8, 8--12, and 12--16 Hz (hereafter referred to by central frequencies, i.e., 6, 10, and 14 Hz, respectively). At this stage, we only exclude events located farther than twice aperture distance beyond which MFP cannot reliably resolve location. This prefiltering step has a negligible effect on the overall statistics, as most of the sources are located close to the array centre.\\
            
\begin{figure}
                        \centering
                        \includegraphics[width=1\linewidth]{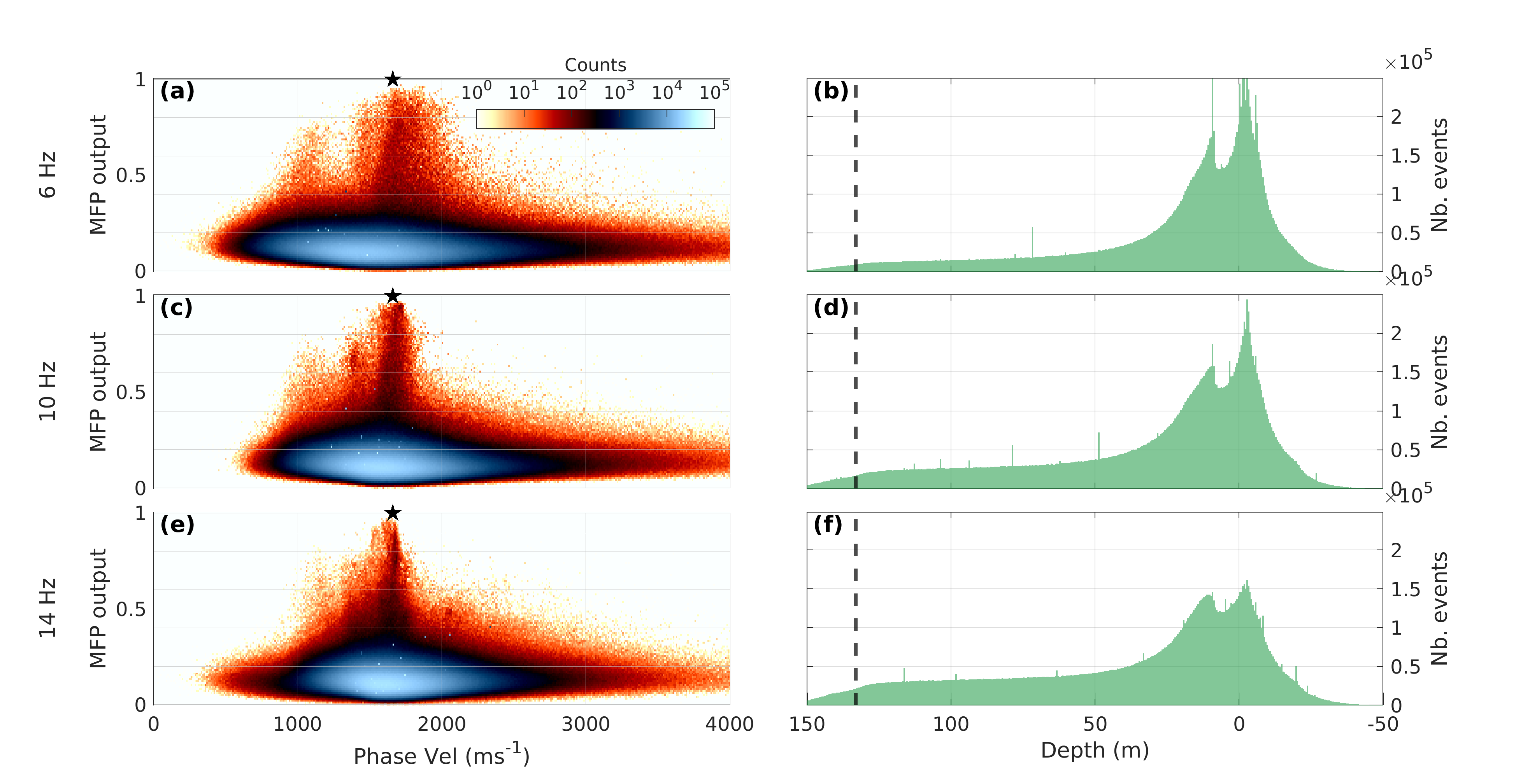}
                        \caption{Statistics of MFP results parameters for selected frequency bands with central frequencies of 6, 10, and 14 Hz. The left column shows 2D histograms of MFP output vs. phase velocity. Black star marks the starting velocity (1660 m/s). The right column shows histograms of the resulting depth. Dashed line marks the upper boundary of starting points depth, defined at constant elevation 10 m a.s.l., equivalent to depths from 192 to 133 m.}
                        \label{fig:stats}
                    \end{figure}

            Fig. \ref{fig:stats} presents statistical characteristics of MFP-imaged sources within these bands. Most sources exhibit low-MFP output values (below 0.35) testifying for low signal phase coherence. Distinct peaks in high-MFP output, representing highly coherent waveforms, correspond to phase velocities between 1600 and 1700 m/s across all frequency bands. These align with expected Rayleigh wave velocity range derived from observed dispersion curve and hammer-shots moveout analysis (Supporting Information, Fig. S3). Rayleigh velocity peaks become sharper with increasing frequency, reflecting a frequency-dependent wavelength effect. A less pronounced peak is observed near 1150 m/s, however, it diminishes at higher frequencies. The depth distributions show minimal variation between different frequencies and reveal that most of the sources cluster near the surface, up to 30 m deep. Importantly, the 2D geometry of the array imposes large vertical uncertainties, hence, the exact depths of particular events should be treated with caution. Source depth distribution is independent from starting points depth positioned at an elevation of 10 m a.s.l. (over 130 m deep) (Fig. \ref{fig:stats}b,d,f). 

\begin{figure}
                             \centering
                             \includegraphics[width=1\linewidth]{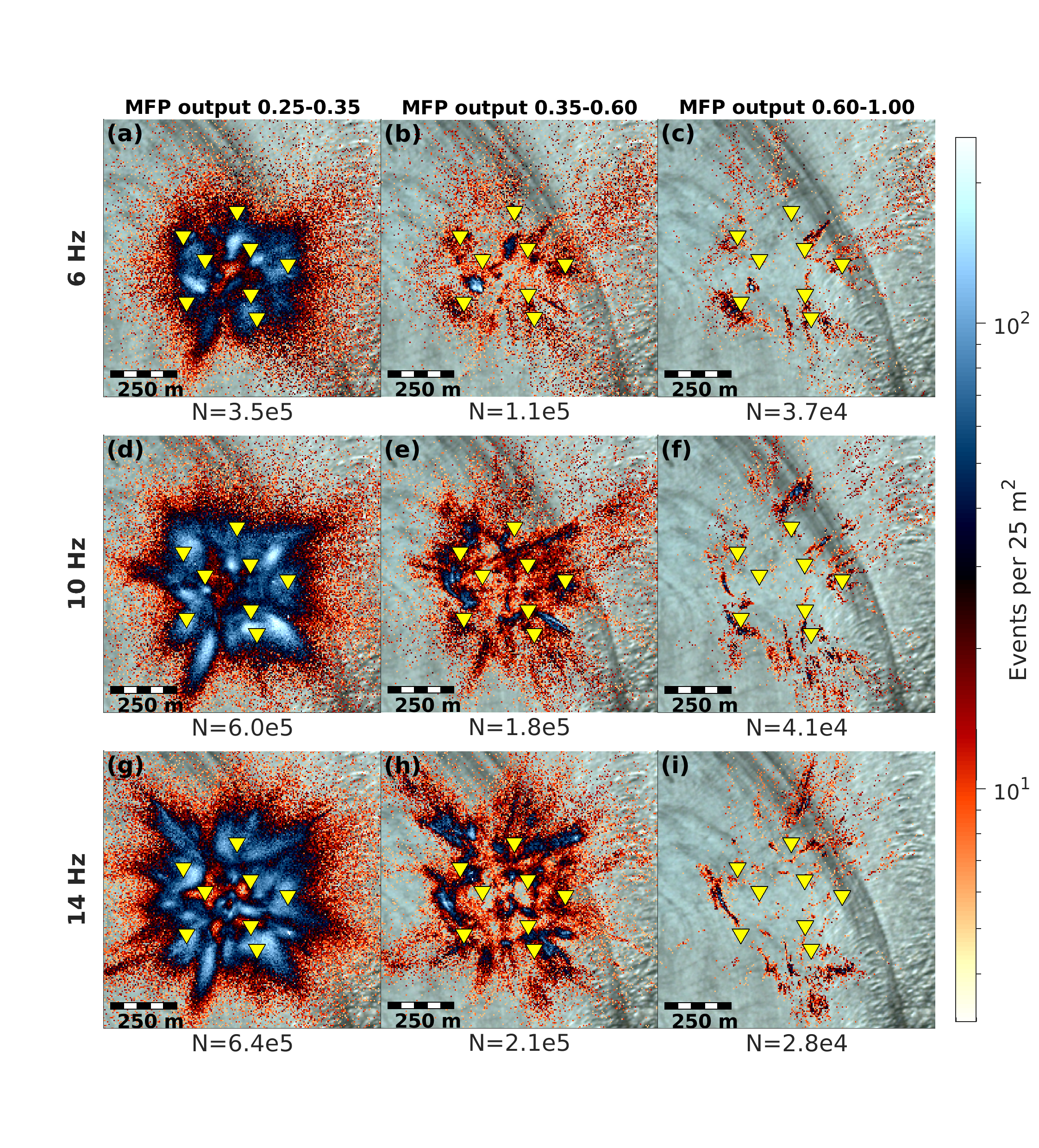}
                             \caption{Map views show binned source locations obtained with 5×5 m bins plotted on top of a satellite photo. Bins with fewer than 3 events are transparent. Low MFP output in left column: $\in \langle 0.25, 0.35 \rangle$, medium in the center: $\in \langle 0.35, 0.6 \rangle$, high in right column: $\in \langle 0.6, 1 \rangle$. Frequency bands with central frequencies of 6, 10, and 14 Hz according to row labels. Lower MFP output ranges are presented in the Supporting Information (Fig. S4). Triangles mark seismic station locations. The N number indicates the total number of events in each panel.}
                             \label{fig:freq_comp_new}
                         \end{figure}

        Next, we map the MFP sources to examine their spatial distributions and explore the differences in the specific MFP output ranges and frequency bands. To enable interpretation, we first apply general filters to the results shown in Fig. \ref{fig:stats} to remove unrealistic solutions. The filters preserve sources with phase velocity between 500 and 3000 m/s, and depth within 150\,m to -15\,m. We include sources located up to 15 m above the expected elevation, accounting for both vertical location uncertainty and imprecision of elevation data. Then we map all the remaining results within low (0.25--0.35), medium (0.35--0.6), and high (0.6--1.0) MFP output bands (Fig. \ref{fig:freq_comp_new}). Sources with very low coherence in MFP ranges below 0.25 do not provide interpretable results (Supporting Information, Fig. S4).

Low MFP output results represent locations of sources with low coherence. This MFP band yields locations that are poorly constrained but correspond to the most active areas visible in the medium MFP results. We are unable to see any meltwater channel signatures regardless of increased modeled runoff (Fig. \ref{fig:map}f). 
The spatial and temporal patterns in the medium-MFP results differ from the highest band, suggesting different source characteristics. Nevertheless, we can track some of the same features in both medium- and high-MFP results.      
The high-MFP distributions remain relatively isolated and consistently shaped across frequencies, with linearity increasing with frequency. In this band, we note increased activity during the warmer half of the day.
Conversely, the medium-MFP results are evenly distributed during the day and differ substantially among different frequencies, especially for the 6 Hz case (Fig. \ref{fig:freq_comp_new}b), where a compact point source dominates.  Therefore, in the next section, we analyze this distinct point source separately from the temporal variability of medium- and high-MFP results at 10 Hz, which we find representative for seismicity patterns observed in both higher-frequency bands.

\begin{figure}
    \centering
    \includegraphics[width=1\linewidth]{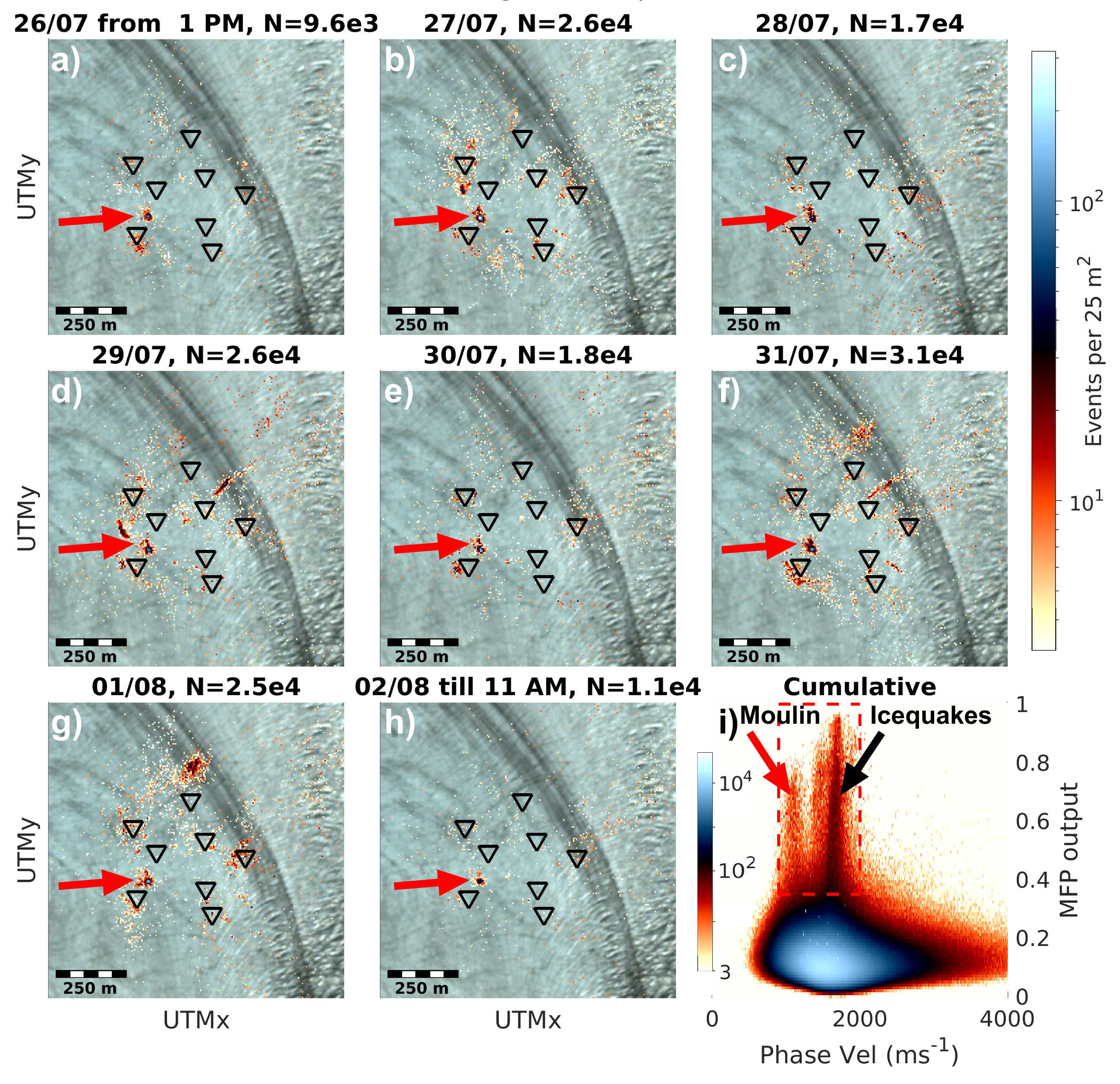}
    \caption{
        Time-lapse imaging of noise sources at 6 Hz. Map views in panels (a)--(h) present locations of noise sources with MFP$>$0.35 for each day; red arrows point to the stable noise source identified as a moulin. Panel (i) shows a 2D histogram of phase velocity vs. MFP output. The red box indicates filtering limits applied to the presented results. Red and black arrows show velocities characteristic of moulins and icequakes, respectively. Triangles mark seismic station locations. Bins with fewer than 3 events are transparent. The top-right colorbar applies to panels (a)--(h). 
    }
    \label{fig:daily_moulin_6}
\end{figure}

\begin{figure}
    \centering
    \includegraphics[width=1\linewidth]{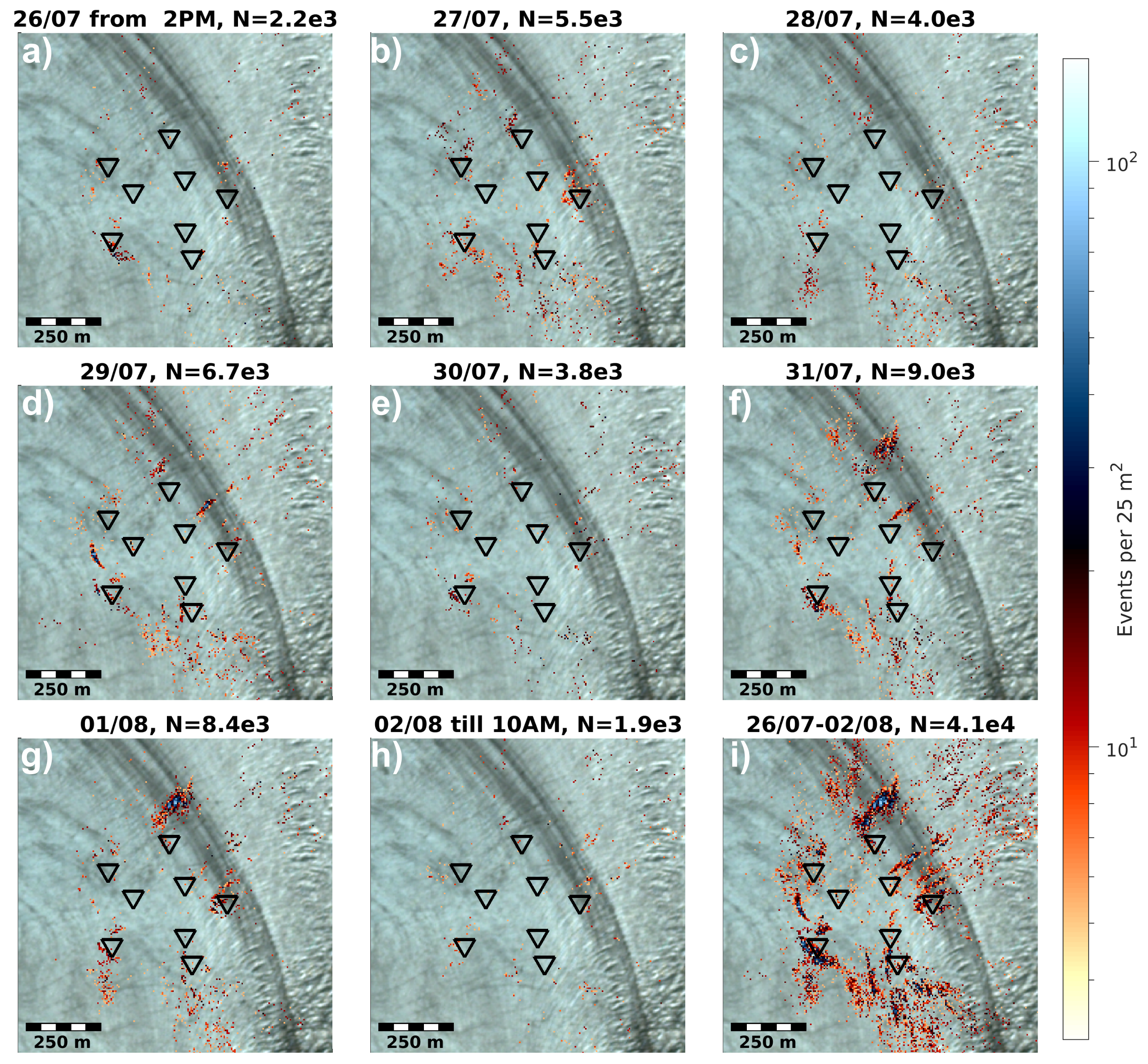}
    \caption{ Time-lapse imaging of high-MFP ($>$0.6) noise sources at 10 Hz. Map views in panels (a)–(h) present locations of noise sources for each day, and panel (i) shows a cumulative distribution. Triangles show locations of seismic stations. Sources are binned in 5x5m bins; bins with counts lower than 3 are masked.}
    \label{fig:daily_high_8hz}
\end{figure}

 When examining temporal evolution in the 6 Hz data (Fig. \ref{fig:daily_moulin_6}, combining medium- and high-MFP), the dominant point source persists consistently throughout the observation period. In contrast, other regions show strong day-to-day variability. The persistent source is associated with velocities around 1150 m/s, while other sources correspond mostly to Rayleigh-wave velocities and sometimes exhibit elongated shapes (e.g., on 31/07 in Fig. \ref{fig:daily_moulin_6}f). We interpret the stable point source as a moulin, surrounded by more transient, spatially variable icequakes.
 The temporal distribution of 10 Hz high-MFP results (Fig. \ref{fig:daily_high_8hz}) changes daily, indicating time-limited activity of the sources in particular areas, however, some clusters are active repeatedly. The well-focused signatures indicate repeating sources in particular locations. These sources are predominantly aligned spatially in linear clusters, their strikes are aligned locally but vary over the whole area. Clusters located in the north-eastern quarter strike towards $38^\circ$N, while those in the southern region strike towards $166^\circ$N. Both strikes align with glacier surface patterns visible in the satellite photo.
 
        \subsection{Surface Icequakes Relocation}
        MFP successfully reveals patterns in the distribution of noise sources; however, the method’s accuracy and applicable signal frequencies are fundamentally constrained by the wavelength. Thereby, the radial resolution is insufficient to reliably associate icequakes with individual crevasses.
        To increase the accuracy of icequakes' locations and obtain a more detailed insight into their spatio-temporal distribution, we apply source position inversion based on discrete differential traveltimes, adapting the method of \citeA{roux2010observation}. 
        Surface icequakes are most commonly characterised by impulsive signals dominated by Rayleigh waves with preceding weak P‐waves \cite{neave1970icequakes, walter2009moment, mikesell2012monitoring, lindner2019crevasse}. To select suitable events, we scan the MFP time series looking for highly coherent impulsive arrivals and select all 2-s-long periods with MFP output greater than 0.5, out of which the majority are expected to be surface icequakes. 
        To compute time delays, we apply a two-pass band-pass second-order Butterworth filter between 5 Hz and 40 Hz and normalise the waveforms by their peak-to-peak amplitude. Then, we extract the Rayleigh phase within a 1 s window centred on the median position of the maxima across all traces. 
        Next, we apply time-domain trace cross‐correlation to pick the arrival time delays of the dominant Rayleigh waves for each station pair, accepting only those with a maximum correlation coefficient exceeding 0.6 to ensure reliable picks \cite{roux2010observation, walter2015green}. Exemplary icequake records and associated correlation coefficients are presented in Fig. S2 in the Supporting Information. 
        Subsequently, we solve for hypocentral coordinates iteratively for velocities in the range 1300–1900 m/s. Such a procedure enables us to infer the best-matching velocity by tracking the median misfit of events. Eventually, a velocity of 1665 m/s yields the smallest median misfit.

        Using the best-matching velocity, we successfully invert source locations for 34,714 events out of 44,745 time windows with MFP $>$ 0.5. Next, to ensure reliable interpretation, we exclude poorly resolved results, retaining 22,467 with location residuals lower than 6 ms (1.5 samples at 250 Hz). This threshold is chosen arbitrarily based on the distribution of residuals and represents a strict selection criterion. We retain 19,913 events that are spatially fitting in the area of interest. Moreover, to focus exclusively on the near-surface icequakes activity, we exclude events located further than 40 m vertically from the expected ice surface, i.e., about half of the wavelength (37 m is half of the expected wavelength for a velocity of 1665 m/s and central frequency of 22.5 Hz). Finally, a total of 10,860 events are used for the surface seismicity interpretation. 
        The resulting catalog is characterized by a median $M_S$ of -1, a mean $M_S$ of -0.88, a b‑value of 0.82, and a magnitude of completeness of –1. The magnitude–frequency distribution is provided in the Supporting Information, Fig. S5.
        
\begin{figure}
    \centering
    \includegraphics[width=1\linewidth]{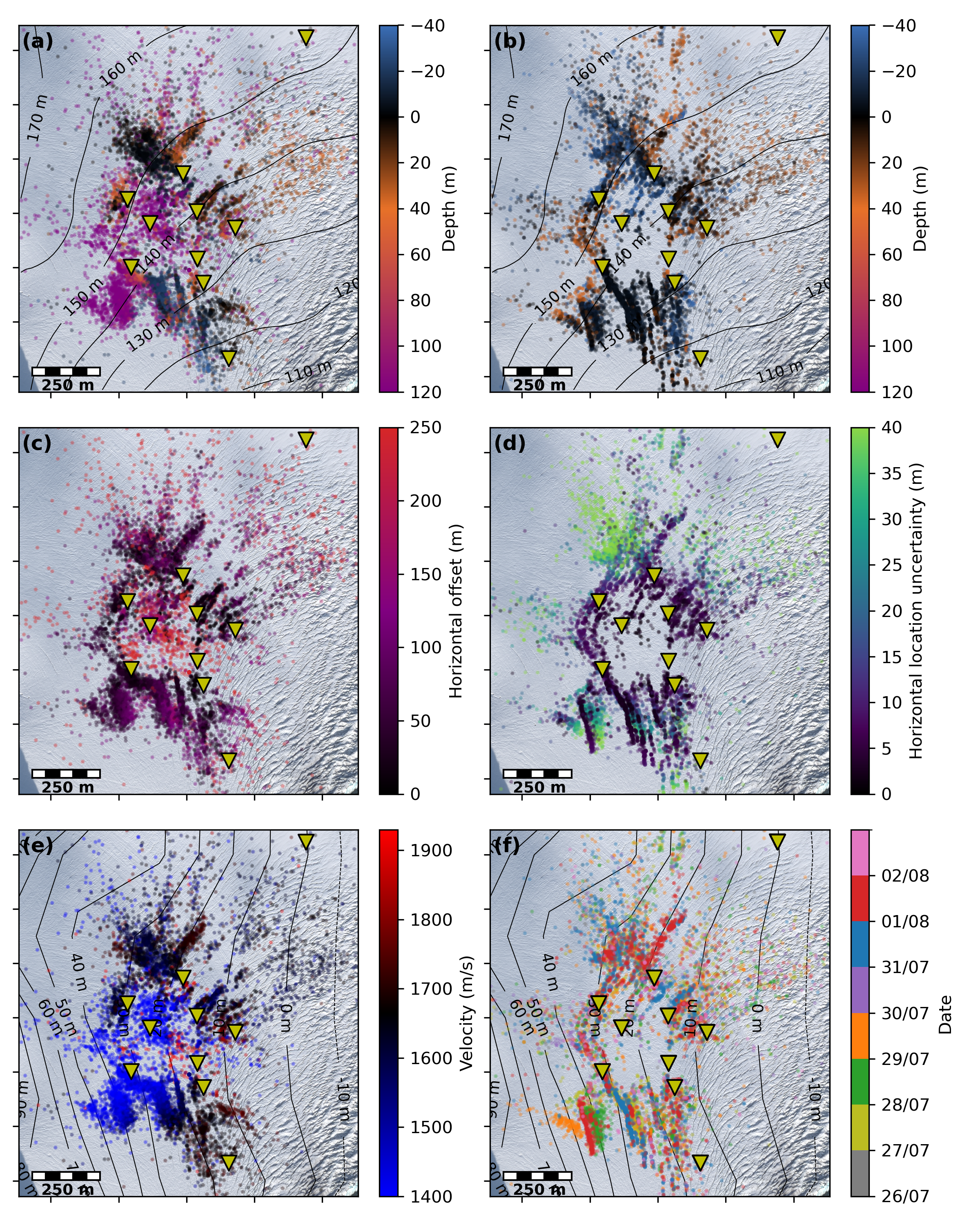} 
    \caption{Map views of MFP and relocated events. The left column shows MFP locations, and the right column shows relocated events. Triangles mark stations. (a) depth-coded MFP locations and surface elevation contours; (b) depth-coded relocated events and elevation contours; (c) MFP locations colored by horizontal difference between original and relocated position; (d) relocated sources colored by horizontal  1-sigma location uncertainty; (e) MFP locations colored by inverted velocity, color scale is centred around the relocation velocity 1665 m/s; (f) relocated events colored by origin time and bed elevation contours. Color scales in panels (c)-(e) were clipped.
}
    \label{fig:relocated}
\end{figure}

        We compare the relocated results against the original MFP output in Fig. \ref{fig:relocated}. After relocation,  most of the deeper events shift towards the glacier surface (Fig. \ref{fig:relocated}a--b). The spatial distribution of relocated events resembles the MFP locations within the seismic array (mostly in the northern half of the analysed area), but differs significantly in the southern part, where sources are outside of the array. Primarily, the spatial resolution was increased, and the MFP elongated clouds were refined to linear features that follow the fractures on the ice surface visible in a 0.5 m resolution satellite image (taken on 09/09/2023, Fig. \ref{fig:relocated}). From the spatio-temporal distribution (Fig. \ref{fig:relocated}f), we infer icequakes associated with individual crevasses that are active at specific times. The resulting horizontal offset between MFP and relocated sources indicates that main features of distribution were preserved (Fig. \ref{fig:relocated}c). Horizontal location uncertainty for relocated events remains below 10 m for the main network of crevasses (in Fig. \ref{fig:relocated}d). Spatial modeling of horizontal and vertical location error is presented in Supporting Information (Fig. S6).  When solving simultaneously for depth and velocity (the MFP case), especially with 2D array geometry, a depth-velocity trade-off is expected. This is visible in the MFP results where deep locations are characterised by low velocity (Fig. \ref{fig:relocated}a,e), resulting in a similar slowness as expected for higher-velocity arrivals from near-surface sources. Fixing the velocity during relocation eliminates the trade-off and provides a more reliable estimate of the source depth.

       \subsection{Crevasse Propagation}
\begin{figure}
    \centering
    \includegraphics[width=1\linewidth]{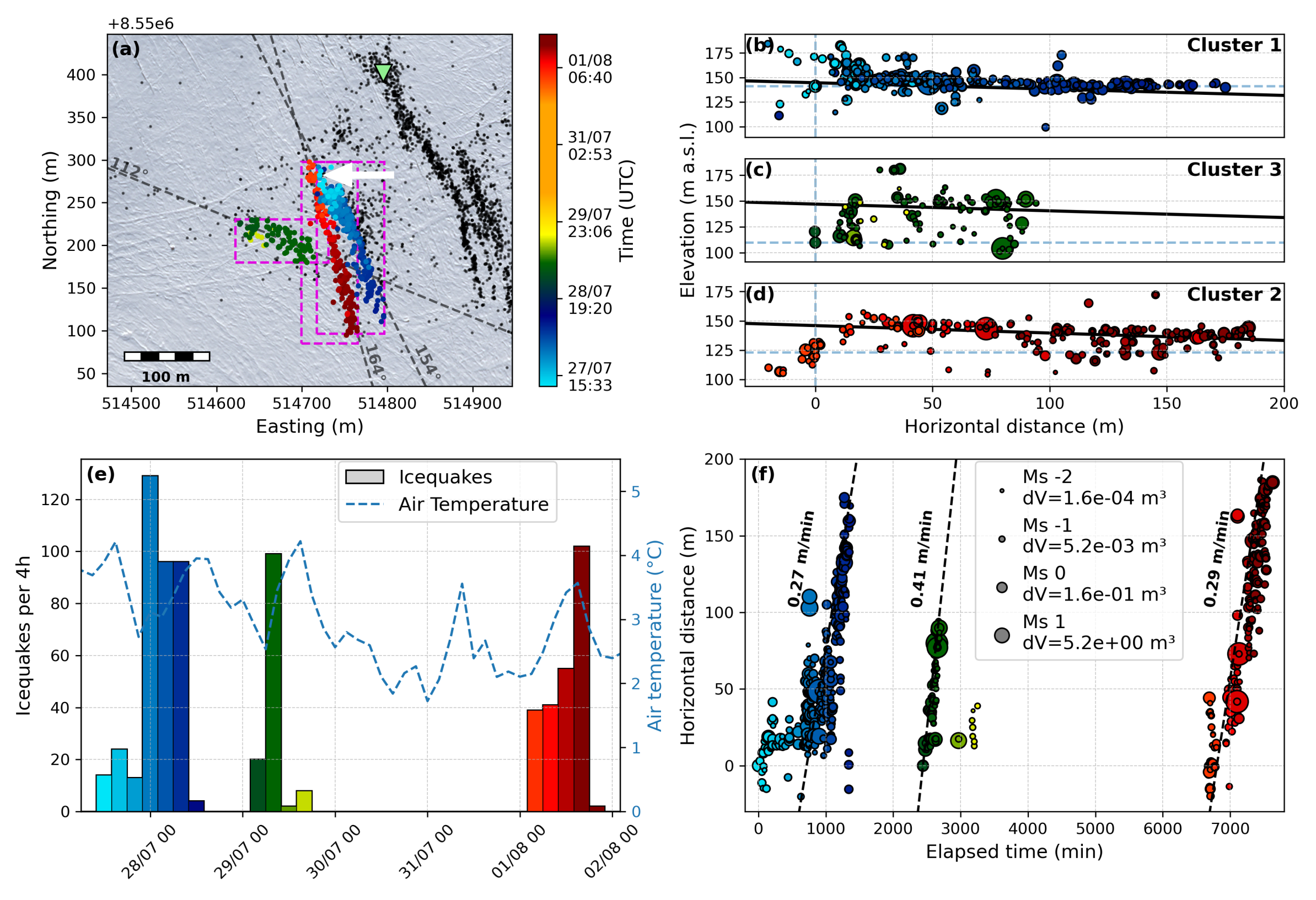}
    \caption{Spatio-temporal evolution of surface icequake clusters. (a) A zoomed-in map view showing icequakes selected for analysis (determined by pink rectangles), colored by time of occurrence (colorbar applies to all panels), and other relocated icequakes (black dots). The green triangle represents the geophone location. The dashed grey line represents the azimuth of each cluster axis. The white arrow shows the location of the crevasse identified in the field in 2023, shown in Fig. \ref{fig:sup_crevasses_photos}. (b-d) Elevation vs distance of each cluster colored by time. Their size indicates individual $M_s$, and the corresponding value of dV is provided in the legend box. The grey solid line shows the approximated elevation of the glacier surface, and blue dashed lines cross at the location of the first event in each sequence. (e) Histogram of selected icequakes occurrence time in 4-h-long bins colored by occurrence time. The blue line shows air temperature from CARRA, no precipitation was recorded at that time. (f) Crossplot showing horizontal distance between the first and subsequent events vs time of occurrence scaled by $M_s$.}
    \label{fig:crevasse_diagnostic}
\end{figure}

\begin{figure}
    \centering
    \includegraphics[width=1\linewidth]{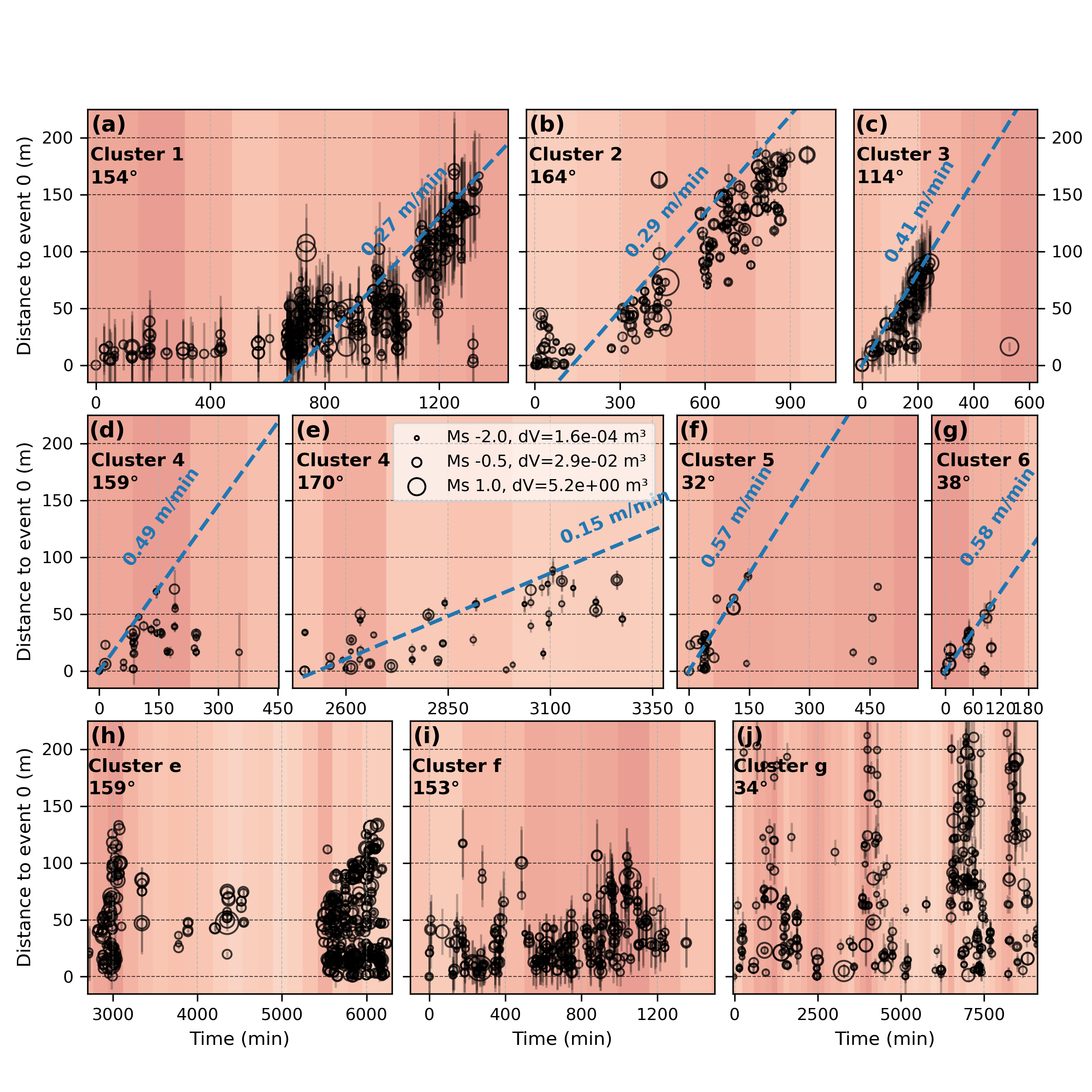}
    \caption{Estimation of crevasse propagation rates assuming linear propagation (blue dashed lines). Horizontal distance between the first and subsequent events vs time of occurrence after the first event, scaled by $M_s$ (black circles) extracted for selected clusters. Black vertical bars represent horizontal location uncertainty. The top row contains well-resolved and time-constrained clusters, the middle row contains time-constrained clusters with a low number of icequakes, and the bottom row shows clusters characterized by icequake activity that is not constrained in time. Time-distance ratios are constant row-wise, except for the bottom row. The strike of each cluster is given below the panel label (see Supporting Information, Fig. S7 for the respective locations).
}
    \label{fig:crevasse_rates}
\end{figure}

       We examine in detail 14 manually selected prominent clusters of relocated icequakes—reflecting all available cases, due to overlapping or distributed events in others—to better understand the spatiotemporal evolution of crevasses. Among these, three co-located clusters exhibit evident and well-constrained in time propagation of icequakes along clusters axes (Fig.~\ref{fig:crevasse_diagnostic}), indicating fracture opening or growth; another three show distributed seismicity with occasional propagation within short episodes, although limited event counts may bias rate estimates; eight display spatially distributed episodic activity without clear directivity, most likely reflecting stress release along existing crevasses. Fig. \ref{fig:crevasse_rates} compares different classes of analysed clusters, while their locations are presented in the Supporting Information (Fig. S7).

        Using the three well-constrained clusters and assuming linear propagation, we estimate the propagation rates between 0.27 and 0.41 m/min. These event sequences last between 4 and 12 h and propagate up to 190 m. We find no concurrent environmental forcing (e.g., consistent temperature change or precipitation) that triggers the propagation (Fig.~\ref{fig:crevasse_diagnostic}c). The other three clusters yield propagation rates of the same order of magnitude (0.15--0.57 m/min, Fig. \ref{fig:crevasse_rates}d--g), however, due to the low number of samples, we restrict further interpretation to well-resolved clusters.

\section{Discussion}

We analysed 8 days of passive cryoseismological records. Our analysis determined the local thickness of the glacier, identified persistent and transient noise sources, and allowed for a successful relocation of surface icequakes, documenting an active crevasse propagation sequence. 

        We inverted local glacier thickness of 120 m, corresponding to a bed elevation of 18 m a.s.l., which aligns closely with the average bed elevation beneath the array of 19.1 m a.s.l. (max: 32.4 m, min: 6.2 m, std: 7.8 m) derived from a Digital Elevation Model based on GPR and GPS measurements from 2005 and 2008 \cite{grabiec2012hanssurface}. 

        \subsection{MFP for Noise Sources Characterization}
        We successfully applied MFP methodology to a dataset from an 8-element, 417 m aperture array, enabling analysis of the spatio-temporal distribution of noise sources among various frequency bands. The obtained results reveal spatio-temporal patterns of glacial seismicity varying with frequency content (Fig. \ref{fig:freq_comp_new}). MFP distinguished low-velocity (ca. 1150 m/s) and low-frequency (4–8 Hz) sources, which we interpret as surface meltwater transport, and broader frequency range icequakes with velocities centred around 1660 m/s expected for Rayleigh wave, aligning with existing observations \cite{walter2015green, nanni2022dynamic}. 
        
        Together, moderately and highly coherent results at 8--12 Hz (Fig. \ref{fig:freq_comp_new}e,f), reveal structurally controlled patterns of sources (likely crevasse‐driven) amid a more scattered field of moderately coherent activity that is not aligned with surface crevasses, perhaps distributed microfracturing or meltwater flow. Strong meltwater sources, such as a moulin, are characterised best by lower (4-8 Hz) frequencies, as in the case of the single source being stable through the whole observation period (Fig. \ref{fig:freq_comp_new}b). By contrast, in the high-MFP output, we observe swarms of icequakes reoccurring at several sites over time (Fig. \ref{fig:daily_high_8hz}). 
        A partially different pattern is observed among medium-MFP, higher-frequency sources (Fig. \ref{fig:freq_comp_new}h) that are reoccurring similarly to the icequakes. In some instances, these may be very weak icequakes seen only by a part of stations, however, in some areas, their strike differs from the strikes visible in the high-MFP source clusters, suggesting a different signal origin.
        Due to the limited vertical aperture of 27 m, the 2D array geometry does not allow for reliable estimation of the source depths. Nevertheless, it is possible to affiliate most of the activity with surface or shallow processes based on the depth statistics (Fig. \ref{fig:stats}c). 
        Unlike \citeA{nanni2021observing}, we don't observe signatures of subglacial water flow, despite the presence of mapped meltwater channels beneath the array location \cite{grabiec2012hanssurface,decaux2019waterpaths} and daily variations in runoff (Fig. \ref{fig:map}f). This absence of seismic signatures could be due to the relatively small aperture of the seismic array, which limits the sensitivity to seismic signals with distinct phase signatures. Additionally, crevasses located within the seismic array—unlike those surrounding the array in \cite{nanni2021observing}— might limit the sensitivity to weaker seismic noise sources, and thus to subglacial meltwater. Finally, we suggest that the relatively dry surface conditions (Fig. \ref{fig:sup_crevasses_photos}) promote low meltwater input to the glacier bed, and, thus, turbulent water flow within the subglacial drainage system.

        \subsection{Relocation of the Surface Seismicity}
       The relocation procedure improves the resolution of the surface icequakes thanks to the higher-frequency content of the surface wave, which exceeded the capabilities of MFP for the given array geometry. The two auxiliary stations further increased the resolution since 29/07, resulting in median horizontal location uncertainty dropping from 15 to 7 m. This effect was not observed for the median vertical uncertainty, which remained around 18 m due to the limited vertical network aperture of 27 m. Nevertheless, network geometry together with quality control criteria resulted in a spatial resolution allowing for detailed analysis of crevasse propagation in the area of interest (Fig. \ref{fig:relocated}d). 
       
       To obtain the most accurate velocity estimate, we iteratively relocate the icequakes over a range of velocities, ultimately identifying 1665 m/s as the value minimising the median of location residuals. This approach ensures the chosen velocity is optimal for the largest number of events.
       The optimal velocity (1665 m/s) is typical for Rayleigh waves in ice and consistent with our independent estimates: 1671 m/s (median high-MFP - Fig. \ref{fig:stats}), 1705 m/s (surface wave dispersion analysis: Supporting Information, Fig. S8), 1601 m/s (hammer shots:  Supporting Information, Fig. S3), and 1665 m/s (inversion following \citeA{roux2010observation} with depth constrained to surface). Of these, the 1671 and 1665 m/s values are most reliable as they were derived from the same waveforms. The latter result also confirms the surface origin of the majority of icequakes, as inversion following \citeA{roux2010observation} assumes depth equal to 0. Velocities derived from hammer shots are prone to bias due to possible wave conversions at greater offsets that control the linear fitting, whilst the dispersion analysis recovers the S‑wave velocity, which is then converted to the Rayleigh‐wave velocity. 
    Our estimates are consistent with those derived from active-source experiments conducted at a now-melted site located approximately 1.5~km south of our study area: \(1600 \pm 50~\mathrm{m/s}\) reported by \citeA{cichowicz1983icequakes} and \(1637 \pm 25~\mathrm{m/s}\) by \citeA{gorski2014seismic}.
    Even though the Rayleigh wave velocity may vary across the study area due to factors such as crevasse-induced anisotropy \cite{sergeant2020green}, using fixed, data-informed velocity provides more reliable source position estimates than inverting for both depth and velocity. The depth–velocity trade-off is removed, leaving location inaccuracy primarily dependent on velocity estimation errors, picking precision, and array geometry.

    The Monte Carlo method of uncertainty estimation that we apply to each event captures the effects of array geometry, along with observational and velocity-related factors, on the final location precision. Uncertainties are based on individual residual misfits. As residual misfits between modeled and observed differential arrival times depend on the velocity model accuracy, we include the velocity uncertainty within data-driven bounds. Our approach allows us to study both the expected spatial uncertainty, which depends on array geometry, and event-specific location accuracy, which depends on arrival-time picking and velocity. Moreover, we may model the expected location uncertainty distribution at any depth for a given residual misfit. Fig. S6 (Supporting Information,) presents horizontal and vertical location uncertainties expected on the approximate glacier surface with a residual misfit of 6 ms, equal to the chosen upper cutoff and, hence, the maximum tolerated uncertainty.

    \subsection{Crevasse Orientation and Propagation}
    
Surface icequakes form elongated clusters whose strike coincides with crevasse patterns visible in satellite imagery (Fig. \ref{fig:relocated}). Due to the limited resolution of the 2021 imagery, during detailed interpretation, we rely on Planet SkySat imagery from 09/09/2023 (0.5 m resolution). This dataset provides detailed constraints on surface fracturing and highlights alignment between crevasses and relocated seismicity.  
    
    Based on relocated seismicity, we interpret two dominant strikes of icequake clusters: $42^{\circ}$ (northern region) and $167^{\circ}$ (southern region), closely matching those obtained from MFP analysis ($38^{\circ}$ and $166^{\circ}$, respectively). Spatiotemporal distributions of events within three neighbouring clusters of elongated seismicity suggest ongoing rupture propagation, with strikes of $164^{\circ}$, $154^{\circ}$, and $112^{\circ}$. Two strikes are consistent with field observations from 22/08/2023 of narrow and shallow crevasses striking approximately $155^{\circ}$ at the same location (Fig. \ref{fig:sup_crevasses_photos}).    The crevasse orientation shifts from longitudinal on Hansbreen (parallel to the terminus) to latitudinal, parallel to the topographic decline of the Tuvbreen bed (Fig.~\ref{fig:relocated}f). We believe that this pattern reflects a structural transition between the extensional regimes of Hansbreen and Tuvbreen. 

    The temporal continuity and spatial coherence of icequake sequences characterize the three best-documented crevasse propagation episodes. In these co-located cases, seismicity propagates down-glacier toward the terminus, consistent with one of the crevasse opening patterns described by \citeA{neave1970icequakes}. The absence of external triggers—such as precipitation or significant temperature variations—suggests that these events represent a self-sustained rupture process, internally driven by stress redistribution. The estimates of volume change (dV) associated with these events are \(7.9 \times 10^{-3}~\text{m}^{3}\) (mean) and \(5.2 \times 10^{-3}~\text{m}^{3}\) (median). Our estimates are one order of magnitude larger than those reported by \citeA{walter2009moment} and one order of magnitude smaller than the mean estimate of \citeA{mikesell2012monitoring}.    
    The well-constained propagation rates of up to 0.41, 0.27 and 0.29 m/min were sustained over 4, 12 and 11 hours, respectively (Fig. \ref{fig:crevasse_rates}). These values correspond to 0.33 m/min observed on Argentière Glacier \cite{nanni2022dynamic} and to 0.63 m/min reported for Bench Glacier \cite{mikesell2012monitoring}, but are two orders of magnitude lower than those observed at Athabasca Glacier \cite{neave1970icequakes}, all three being land-terminating mountain glaciers. 

  The two orders of magnitude difference in crevasse propagation rate and timescales reported by \citeA{mikesell2012monitoring} and \citeA{neave1970icequakes} measured in comparable glaciological settings was attributed by \citeA{mikesell2012monitoring} to possible discrepancies arising from different location algorithms. However, judging from estimated rates and sequence duration, we interpret the difference as a result of two distinct processes — with \citeA{neave1970icequakes} likely capturing a brittle crevassing event, while \citeA{mikesell2012monitoring} observed creeping propagating with diffusive rates, similar to \citeA{nanni2022dynamic} and our observations. 

\begin{figure}
    \centering
    \includegraphics[width=1\linewidth]{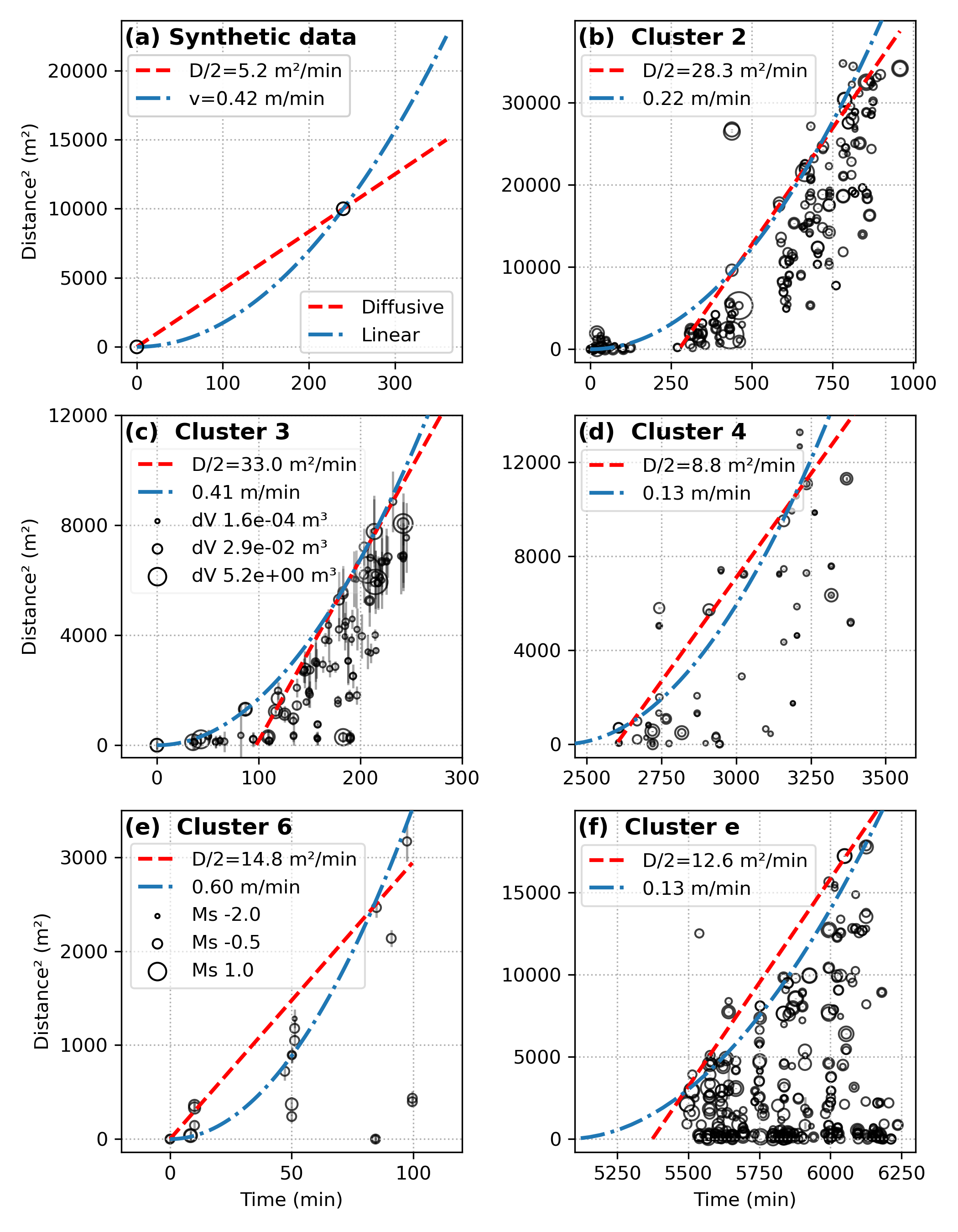}
    \caption{Estimation of crevasse rate propagation with diffusive (squared distance over time, red dashed line) and linear (distance over time, blue dash-dotted curve) propagation models plotted in the squared-distance domain. Panel (a) shows synthetic data. Black circles with error bars represent icequakes with associated horizontal location error scaled by $M_s$ (explained in panel (e)), with equivalent dV values shown in panels (c). Identical symbol sizes are used across legends and apply to all six panels.
    
        }
    \label{fig:diffusive_propagation}
\end{figure}

 Within the framework of fracture mechanics in ice, subcritical creep is expected to produce diffusive propagation ($d^{2} \propto t$), consistent with crack growth initiated at a localized nucleation point and governed by time-dependent stress redistribution \cite{hageman2024ice}. In contrast, brittle fracturing consistent with linear elastic fracture mechanics (LEFM) results in approximately linear propagation ($d \propto t$) due to rapid crack advance under concentrated stress at the crack tip \cite{neave1970icequakes}. However, in natural settings like an active crevasse field, propagation behaviour may deviate from these idealized end-member scalings. Given the subjectivity of interpretation and the small number of well-constrained clusters, we cannot unambiguously distinguish between these mechanisms based on our data, as it can be interpreted using both linear and diffusional rates (Fig. \ref{fig:diffusive_propagation}).
  
    The diffusive model quantifies the diffusion coefficients of 28--33~$\mathrm{m^{2}/min}$ (i.e. 0.47-–0.55~$\mathrm{m^{2}/s}$) over 4--12 h for well-resolved clusters (from $D = d^2/(2t)$ for 1D propagation along the crevasse axis, using $\langle x^2 \rangle = 2Dt$).
  The Maxwell time (the crossover of brittle and viscous ice behaviour \cite{jellinek1956viscoelastic}) below a few hours, providing deviatoric stress at the crack-tip above 0.05 MPa for temperate ice \cite{hageman2024ice}, falls below the observed timescale. 
  Estimated rates together with observed timescales testify for sustained subcritical crack propagation as the driving mechanism.
  To better distinguish between brittle and viscous regimes, it is necessary to unambiguously determine the initiation time of the crevassing, as determining the temporal position of the crack opening has a critical influence on the model fit in our analysis, and to further increase detectability and spatiotemporal resolution of icequakes.

\section{Conclusions}
Our two-step analysis of passive seismic recordings from a small (417 m) aperture temporary array on Hansbreen, Svalbard, demonstrates that sparse instrumentation (median interstation distance of 240 m) can achieve meters-scale resolution mapping of glacial processes when combining Matched Field Processing (MFP) with differential traveltime relocation. MFP resolves frequency-dependent noise sources, distinguishing transient surface icequakes from persistent meltwater moulin emissions. Subsequent automatic relocation of 10,860 surface icequakes using differential travel times and an inverted velocity of 1665 m/s improves spatial resolution and provides location uncertainty estimates. The relocated surface icequakes form elongated clusters that closely align with surface crevasses observed in 0.5 m resolution satellite imagery. Together, our approach demonstrates the potential for automated detection and detailed characterization of glacial seismicity across different frequencies.

The high-resolution relocation of surface sources (horizontal location error < 5 m) reveals three well-constrained surface crevasse propagation episodes with linear propagation rates of 0.27–0.41 m/min sustained over 4–10 hours and up to 190 m in length. This propagation can be quantified equally well assuming viscous propagation that yields diffusion coefficients of 0.47–-0.55~$\mathrm{m^{2}/s}$ and positions observed timescales in proximity to Maxwell time. We interpret the crevassing mechanism as sustained subcritical crack propagation, where viscous stress relaxation governs rates of orders of magnitude below elastic limits. Precise determination of crevasse initiation timing is critical for accurate diffusion coefficient estimation, highlighting enhanced spatiotemporal resolution in cryoseismic event location as a promising path for robust analysis of crevasse propagation dynamics.

For applying MFP in challenging glacial environments, we recommend a minimum of eight stations (a lower number was not tested) forming an array with good angular resolution. The inter-station spacing should be less than half the wavelength; here, the shortest wavelength used was 100 m (1665 m/s at 16 Hz). A range (source-to-array distance) of two to three times the aperture of the array is expected \cite{almendros1999array}, but in our case, with eight stations, we were able to effectively detect and resolve sources at distances up to one and a half times the aperture.
Auxiliary stations can increase relocation accuracy for sources outside the array and extend the effective range. They should be placed at a distance comparable to the array aperture because the contribution of stations located further is little to none.
Whenever possible, stations should be installed in shallow boreholes directly in the ice to increase the signal-to-noise ratio, eliminate snow/firn-related effects, and prevent tilting. However, we used stations deployed directly on the ice surface, which is feasible for limited periods under negligible surface melt conditions.

\section*{Open Research Section}
The MFP codes used to locate seismic sources are described and available via \\
https://lecoinal.gricad-pages.univ-grenoble-alpes.fr/resolve/ (last access: March 9, 2023) associated with \citeA{nanni2022dynamic}. 
The relocation code used in this study is provided at \citeA{Gajek_Relocation_Code_2026} and will also be published on GitHub.
For improving data visualisation, we used the utilities by \citeA{Aguilera_Colormap_Colorbar_Utilities_2014} and \citeA{Kumpulainen_tight_subplot_2016}.

The complete seismic dataset used in this study is publicly available from an online repository \cite{Gajek_On_ice_Passive_Seismological_Dataset_2025}.
The digital elevation model of Svalbard was produced by the Norwegian Polar Institute \cite{NPI_DEM}.
The Hansbreen's bed topography was produced by \citeA{grabiec2012hanssurface}.
The meteorological data is available at  hornsund.igf.edu.pl/index.php/en/wather/.
TopoToolbox \cite{schwanghart2014topotoolbox} 
was used to delineate catchment draining.
Arctic Ocean map in Fig. \ref{fig:map}c was modified from "Arctic Ocean SVG" by Norman Einstein via Wikimedia Commons \\
(https://commons.wikimedia.org/wiki/File:Arctic\_Ocean\_SVG.svg). 
Grammarly and Perplexity AI were used to refine the text and correct grammar, and Perplexity AI was additionally used to assist in debugging plotting scripts.

\section*{Conflict of Interest declaration}
The authors declare there are no conflicts of interest for this manuscript.

\section*{Author Contributions}
WG performed the seismic data analysis and wrote the original manuscript. WG and UG significantly revised the manuscript and interpreted the results. AG and WG contributed to the development of the relocation algorithm. WG, WDH, and DMP acquired seismic data. LSS computed the runoff. WDH acquired funding for field campaign. All authors revised the final manuscript.

\acknowledgments
We thank Pierre-Fran{\c{c}}ois Roux and Andreas Köhler for discussions and code sharing.
We acknowledge funding from the Research Council of Norway, project number 291644, Svalbard Integrated Arctic Earth Observing System–-Knowledge Centre, operational phase, and statutory funds of the Department of Geophysical Imaging, Institute of Geophysics, Polish Academy of Sciences. Ugo Nanni has received support from the Research Council of Norway through MAMMAMIA (grant no. 301837) and SLIDE (no. 337228) projects.
Access to Planet imagery was granted within the Svalbard Integrated Arctic Earth Observing System-–Planet cooperation project. The satellite photos used in this work are: PlanetScope taken on  28/08/2021 and SkySat taken on 09/09/2023.
We thank Byron Spencer for his contribution during data acquisition.
This study utilises meteorological data from the monitoring carried out based on the infrastructure of the Stanisław Siedlecki Polish Polar Station Hornsund in Spitsbergen, operated by the Institute of Geophysics, Polish Academy of Sciences, financed by the Polish Ministry of Science and Higher Education. We also thank the crew of PPS Hornsund for their support during field operations.

\nocite{tarantola2005inverse}
\bibliography{my_lit}

%
%
%
%
%

\end{document}